# Cardiomegaly Detection using Deep Convolutional Neural Network with U-Net

Soham S. Sarpotdar
BITS Pilani
f20180630@goa.bits-pilani.ac.in

*Abstract*— Cardiomegaly is indeed a medical disease in which the heart is enlarged. Cardiomegaly is better to handle if caught early, so early detection is critical. The chest X-ray, being one of the most often used radiography examinations, has been used to detect and visualize abnormalities of human organs for decades. X-ray is also a significant medical diagnosis tool for cardiomegaly. Even for domain experts, distinguishing the many types of diseases from the X-ray is a difficult and time-consuming task. Deep learning models are also most effective when used on huge data sets, yet due to privacy concerns, large datasets are rarely available inside the medical industry. A Deep learning-based customized retrained U-Net model for detecting Cardiomegaly disease is presented in this research. In the training phase, chest X-ray images from the "ChestX-ray8" open source real dataset are used. To reduce computing time, this model performs data preprocessing, picture improvement, image compression, and classification before moving on to the training step. The work used a chest x-ray image dataset to simulate and produced a diagnostic accuracy of 94%, a sensitivity of 96.2 percent, and a specificity of 92.5 percent, which beats prior pre-trained model findings for identifying Cardiomegaly disease.

## I. INTRODUCTION

Clinical X-rays were images used to diagnose particular sensitive parts of the human body, like the bones, chest, teeth, or head. This method has been used by medical practitioners for decades to study and detect fractures or malformations in human organs[1]. This is due to the fact that, in addition to being unobtrusive and cost-effective, X-rays are extremely effective diagnostic tools for detecting pathological alterations. Cavitations, restructurings, infiltrates, sharpened costophrenic angle, and small widely scattered nodules can all be seen on CXR images as chest illnesses. Radiologists can diagnose pleurisy, fluid, pneumonia, bronchitis, infiltrate, nodule, bronchiectasis, pericarditis, pericardial effusion, pneumothorax, fracture, and many other disorders and diseases by evaluating the chest X-ray image.

Radiologists find classifying chest X-ray abnormalities to be a difficult undertaking, hence various methods have been presented to help them do it more accurately[2][3]. Over the years, computer-aided diagnostic (CAD) technologies have indeed been created to obtain valuable information in X-rays to assist doctors in acquiring a quantitative knowledge of the X-ray. These Systems, on either hand, are unlikely to have gained sufficient prominence to make decisions about the types of diseases found within an X-ray. As a result, their job has been reduced to that of visualization functionality that aids doctors in making decisions [4].

Artificial intelligence approaches have been used in a number of studies just on diagnosis of chest disorders. In [1], Chest disorders have been diagnosed using multilayer, probabilistic, learning svm classifiers, and generalized regression neural networks. Neural networks and an artificial immune system were used to diagnose chronic obstructive pulmonary disease and pneumonia. In [5], Lung diseases including such tuberculosis, pneumonia, or lung cancer are diagnosed using chest radiography. Histogram equalization was utilized in image segmentation for preprocessing, and unique deep neural networks models have been used for classification. The study works mentioned above were effective in classifying medical disorders; but, in terms of effectiveness, sensitivity, and specificity, they were not as effective as deep networks. Deep learning-based algorithms have been used to improve image classification accuracy [6] [7]. In performing such tasks, these deep networks demonstrated superhuman accuracy. The success led the researchers to use similar networks on ct scans in disease diagnosis, and the results showed that dnns may extract advantageous features that swiftly identify separate classes of images [8]. The deep neural network is the most widely used machine learning architecture (CNN). Because of its ability to extract different level information from images, CNN has been used to classify various medical images [9].

The following is a summary of the rest of the paper. In the second part, we provide a quick overview of recent work on deep learning applications in medical picture processing. The material & methods were then discussed in part three, which began with an explanation of the "ChestX-ray8" dataset. The outcomes of implementing this method for Cardiomegaly disease identification from CXR pictures are shown in the fourth part. Finally, the report finishes with a discussion on future research.

## II. LITERATURE SURVEY

Previous research has revealed extremely high exposure success degrees, such as in Ref. [10], wherever the authors used an ImageNet-based CNN to identify distinct diseases in CXR pictures and got an 89 percent accuracy rate. In alternative work [11], The scientists offered DualNet, a novel architecture that simultaneously analyzes and front side CXR images In this case, the accuracy was reported to be 91%, however the authors employed a huge set of Imitate information (1000's of images). In Ref.[12], The authors were successful at a 92 percent accuracy rate, but only by employing pre-configured models like ResNet-101.

A team form Taishan Medical University conducted another study[13], created an automatic approach for identifying digital radiography photos of a patient's position or body are using a CNN algorithm They used solely frequency curves classification and gray matching to attain this goal; nonetheless, more than 7000 photos were required to reach the 90 percent prediction accuracy.

Recent evidence suggests otherwise [14] allows you to categorize the ChestX-ray8 dataset's eight illnesses The method, however, is based on complicated and pre-trained models such as ResNet-101 therefore ImageNet, as well as it is unable to locate the disease inside the CXR. Another method for using deep learning to detect CXR diseases was to utilize a pre-configured model for general pictorial recognition [15]. Despite this, In the CXR, the model is limited to disease identification but does not provide for disease localisation.

We are interested in not only recognising Cardiomegaly illness and enhancing detection precision over previous work, but also precisely pinpointing the illness within the CXR and presenting a strategy for training that leverages a modest amount of data. Using the UNet-based Classification technique we developed, we were able to get a high evaluation that is indeed an essential element of between 93 to 94 percent from a small dataset during the training phase. In comparison to the cited works, our method has produced good results so far[10][11], that made use of large, pre-configured modules (VGG16, VGG19 and ResNet). It's the first time a procedure like this has been reported, not just for detecting Cardiomegaly illness but also for pinpointing it within CXR images, to our knowledge.

TABLE I. Comparison Table of the cardiomegaly level screening methods

| Literature | Image Database | Method | Medical Purpose |
|---|---|---|---|
| [16] | NIH CXR Image Database (500 PA CXR Images) | 2D U-Net and U-Net + Dense Conditional Random Field (CRF) | Lung and Heart Segmentation and CTR Estimation |
| [17] | JSRT Database (247 PA CXR Images) and Montgomery dataset (138 PA CXR Images) | Standard U-Net and XLSor Model | Lung and Heart Segmentation and CTR Estimation |
| [18] | JSRT Database (247 PA CXR Images) | X-RayNet #1 and X-ray Net #2 | Lung, Heart, and Clavicle Bones Segmentation and CTR Estimation |
| [19] | Faculty of Medicine Siriraj Hospital (7517 PA CXR Images) | U-Net with VGG-16 encoding | Lung and Heart Boundary Location and CTR Estimation |
| [20] | NIH CXR Image Database (566 PA CXR Images) | VGG-16 Model | Cardiomegaly Classification (Cardiomegaly, No-Cardiomegaly) |
| [21] | ChestX-ray8 Database (1010 PA CXR Images) | U-Net-based CNN Algorithm | Diagnosis and Precise Localization of Cardiomegaly Disease |
| [12] | Indiana Dataset (332 PA CXR Images) | ImageNet (DCN) Binary Classification | Abnormality Detection and Localization |
| [22] | NIH CXR Image Database (1800 PA CXR Images) | CNN-based ResNet and Explainable Feature Map | Cardiomegaly Diagnosis |

## III. Material and Methods

### A. Chest X-Ray 8 Dataset

"ChestX-ray8" is indeed an outpatient settings database of 108,948 frontal CXR pictures from 32,717 distinct people, with images classified for eight different illnesses. Teams from the National Library of Medicine's Medical faculty and Imaging Sciences and the National Institutes of Health's Bethesda headquarters launched "ChestX-ray8." AN ELIMINATOR [23]. This dataset is open to the public for research and testing of different computer-assisted detection approaches.

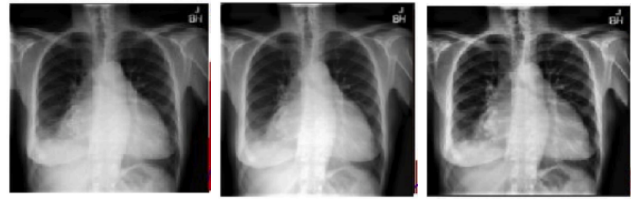

Fig. 1. Sample Images of the Dataset

We extract all Cardiomegaly photos from the Excel sheets provided by "ChestX-ray8" dataset using a primitive Python script that separates files into multiple folders based on their label. We were capable of extracting 1010 Cardiomegaly-related pictures as a result of this. Regardless, these 1024x1024 resolution photos are provided without any masks or comments indicating disease location. As a result, they must be preprocessed before using the Deep Learning approach.

### B. Custom Pretrained Deep Learning Model

We have used a Customized pretrained U-Net model for cardiomegaly detection. In this research, we have performed different operation steps including data collection, data pre-processing, data splitting, feature extraction etc. First, collect the ChestX-ray8 dataset. Apart from this, perform the data preprocessing technique. In the data pre-processing, image enhancement, resizing, and cropping is performed. Contrast adjustment enhancement is frequently employed in image processing; the goal of this technique is to reveal hidden details in the image that may hold relevant information. A block diagram of the proposed approaches is given below:

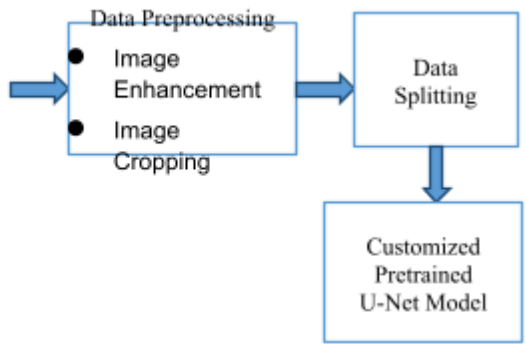

Fig. 2. Block Architecture of the Proposed Approaches

### 1) Convolutional Neural Network (CNN)

In this part, the Deep Neural Network (CNN) and its essential components were discussed. CNN is a powerful visual model of creating intelligent systems that takes any input image and produces a proportionally larger output with much more relevant information. This architecture is built by connecting a group of features using pixel-to-pixel multi-layer integrity, and adding one or more fully linked layers on top. chevalie [24] as seen in Figure 3. The CNN architecture is made up of a variety of successive layers, some of which have been repeated. The most popular layers are described below:

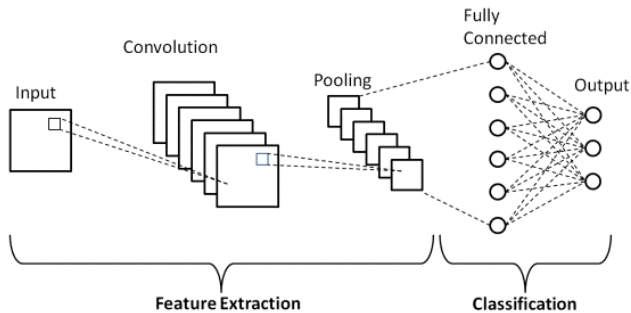

Fig. 3. CNN Architecture

- **Input layer:** provides data entry for numerous photos Using RGB color level representation and conventional measurements (Width x Height).

- **Feature-extraction (learning) sequence:** The method searches for common traits at this level and ranks them in ascending order of relevance. As an illustration of these layers, consider the following:

    a. **Convolution layer:** The most crucial layer in our suggested CNN model is this one, as it is where the majority of computations would take place. The primary function of this layer is to extract characteristics from a picture while maintaining the picture pixels' spatial relationships This is accomplished by applying a series of filters to learn the recovered features.

    b. **Pooling layer:** After a Convolutional Layer, a Pooling Layer is frequently applied. The main purpose of this layer is to shorten the convolution extracted features in order to minimize computing costs. This is achieved by minimizing layer interconnections and operating each feature map separately. Depending on the technique used, there are many types of Pooling procedures.

    The region of interest in Max Pooling yields the largest component. Average Pooling is used to calculate the average of components inside a set size Image segment. The entire sum of the elements in the defined section is calculated using Sum Pooling. The Pooling Layer was commonly used to connect both Convolution operation and FC Layers.

- **Fully-Connected Layer:** The cells in this layer are linked to all of the kernel function from the previous layer. The primary purpose of this layer in this study was to identify the returned convolved characteristics from dataset photos into the appropriate classes.

### 2) U-Net Architecture

Olaf Ronneberger and his colleagues created the UNET architecture[25] for the segmentation of biomedical images There are primarily two paths. The first is an encoder path, whereas the second is a decoding path. The encoder path records the image's context for creating feature maps. The encoder path is nothing more than a stack of convolutional and maximum pooling layers. Using transposed convolutions, a decoder path was employed to provide exact localisation. Because U-net only has Convolutional layers and no Dense layers, it can accept images of any size.

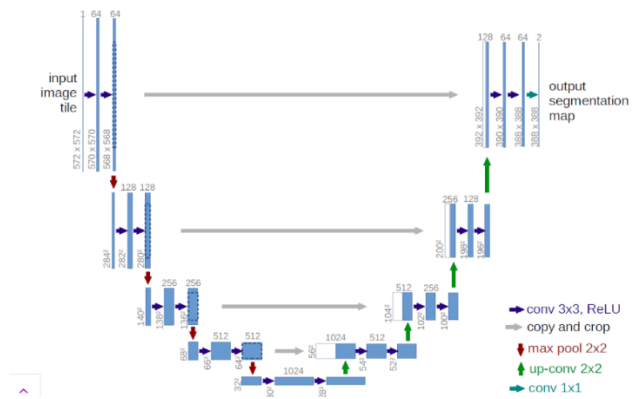

Fig. 4. U-Net Architecture

a) **Contraction/down sampling path (Encoder Path):** It's similar to an encoder that captures context using a compact feature map, and it's made up of four blocks, each having three Convolution Layers. There seems to be an Activation function (having batch normalization) with 2 x 2 Maximum Pooling after each Convolution Layer. The technique doubles the feature map with each pooling, starting with 64 extracted features for the first blocks, 128 for the next, and etc. The input image is the source of this contracting path; the technique retrieves the associated topic in order to partition the image in order to be ready for up-sample via a global feature transformation.

b) **Expansion/Up sampling path (Decoder Path):** Represents the inverse of the previous operation. It acts as a decoder to ensure that the cropped mask is correctly located. It is made up of four blocks, each of which contains a deconvolution layer and a map of cropped attributes from the subsampling stage. The data that was lost during the outsourcing stage's maximum pooling will be rebuilt between these blocks. Another benefit of this technique is that this does not require the use of a dense layer, allowing photos of various sizes to be utilized as input.

## IV. RESULTS AND DISCUSSION

To test & confirm the efficacy of the proposed plan, we chose a distinctive working environment. We chose Kaggle as our data analytics platform since it provided a notebook with open-sourced data, GPU/ CPU options, and a Python-based high-level artificial neural API. Furthermore, being an open-sourced deep learning framework developed for numerical computing, we utilized TensorFlow again for development framework.

### C. Performance parameters

A quality that characterizes a specific element, capability, and attribute of a system and is usually measured by a numerical value. The accuracy, sensitivity, and specificity of a diagnostic test establish its validity, or its capacity to measure what that is designed to assess.

1) **Sensitivity:** Sensitivity is the percentage of real testing positive between all patients with this illness. In other words, the ability of the a test or equipment to produce a favorable outcome for a sick patient The capacity to correctly identify a test is critical, and the sensitivity equation is as follows:

$$Sensitivity = TP/(TP + FN)$$
$$= \frac{Number\ of\ true\ positive\ assessment}{(Number\ of\ all\ positive\ assessment)}$$

2) **Specificity:** The percentage of negative cases among all subjects that Specificity refers to the absence of a disease or condition. To put it another way, it refers to the ability of a test or equipment to give normal and negative results inside the absence of illness. The following is the formula for determining specificity:

$$Specificity = \frac{TN}{TN + FP}$$
$$= \frac{(Number\ of\ true\ negative\ assessment)}{(Number\ of\ all\ negative\ assessment}$$

3) **Accuracy:** Accuracy refers to how well A diagnostic test accurately diagnoses or rules out a problem. A diagnostic test's sensitivity and specificity, as well as the existence of prevalence, can be utilized to determine its accuracy.

$$Accuracy = (TN + TP)/(TN + TP + FN + FP)$$
$$= \frac{Number\ of\ correct\ assessments}{(Number\ of\ all\ assessments)}$$

### D. Results

This section summarizes the results of the suggested U-net model, which was implemented with Python programming.

TABLE II. TRAINING/VALIDATION ACCURACY VERSUS NUMBER OF EPOCHS

| Epochs | Training Accuracy | Validation Accuracy |
|---|---|---|
| 0 | 85.1 | 85 |
| 5 | 87.3 | 85 |
| 10 | 91.3 | 89.9 |
| 15 | 92 | 93.2 |
| 20 | 93.4 | 93.9 |

Table 2 shows the training and testing accuracy as a function of the number of epochs. The training accuracy or validation rises as the amount of epochs employed grows, as shown in this table. We reran the phase without the Early Stopping method and fixed the number from used epochs to 30, 50, 75, and 100 to ensure the accuracy obtained was the best the system could accomplish.

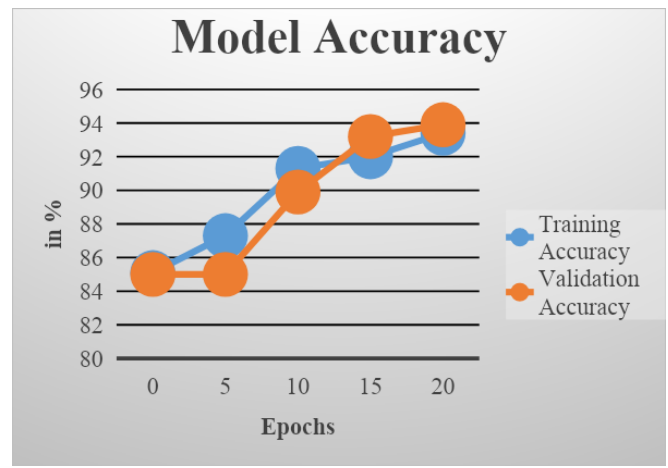

Fig. 5. Model accuracy of proposed model

The training/validation accuracy is plotted against the number of epochs in Fig. 5. The calculations were started by setting the number of iterations to 20, the number of iterations to 20, and the batch size to 32. From the sixth epoch, we see a significant rise in accuracy for both procedures, with values around 90%. The validation procedure' accuracy improves after the tenth epoch, reaching values of 93.5 percent to 94 percent by the fifteenth epoch.

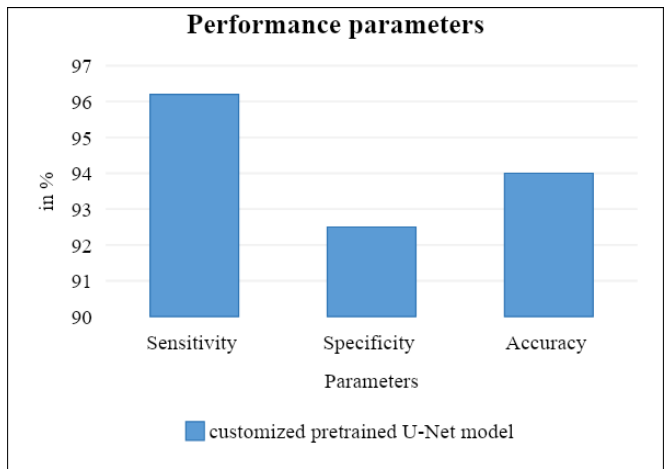

Fig. 6. Results of proposed model

TABLE III. Summary of Overall Performance of The Tested Networks

| Model | Sensitivity (%) | Specificity (%) | Accuracy (%) |
|---|---|---|---|
| Inception V3 [26] | 64.1 | 77.1 | 71 |
| VGG-16[26] | 81 | 81.1 | 81.3 |
| VGG-19[26] | 84 | 83 | 84.5 |
| SqueezeNet[26] | 82 | 80 | 81 |
| Segmentation-based method 1[17] | 95 | 87 | 91 |
| Segmentation-based method 2[17] | 94 | 88 | 91 |
| CardioXNet [27] | 89.29 | 100 | 93.75 |
| Dense-Net [27] | 95.65 | 81.48 | 87.50 |
| ResNet-101[28] | 93 | 91 | 92 |
| customized pretrained U-Net model | 96.2 | 92.5 | 94 |

Figure 6 displayed the performance parameters achieved by the proposed customized pretrained U-net model of CNN. Here x-axis represents the parameters like sensitivity, specificity and accuracy whereas y-axis represents the performance in %. From this figure we can see that the achieved sensitivity is very high (more than 96%) as compare of specificity (less than 93%) and the accuracy results are satisfactory (94%).

*E. Comparison Results Discussion*

In this section, the comparative results of existing pre-trained deep learning models with proposed customized pretrained U-Net models are displayed and discussed.

Table 3 The performance of Google's Inception V3, VGG16, VGG19, Squeeze-Net neural networks, Segmentation-based technique 1, Segmentation-based technique 2, Cardio-X-Net, Dense-Net, ResNet-101 was compared and evaluated with a proposed customized pertained U-Net model utilizing the test set's chest X-rays.

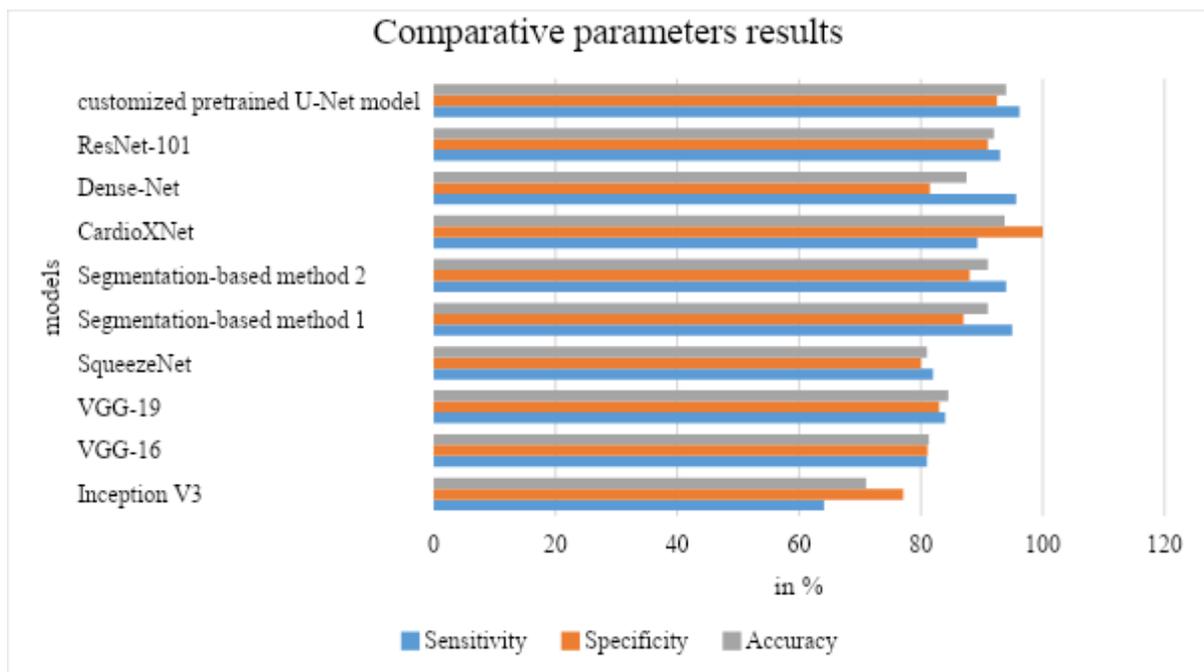

Fig. 7. Comparative performance results

Figure 7 illustrates a comparison of results for three different performance characteristics for various pre-trained CNN models. The specificity from each of the test network was measured, and it was discovered that the specificities were more similar than the sensitivity, indicating that the networks were more homogeneous. However, Google's

Inception V3 network once again had the lowest specificity rate, with a rate of 77.1 percent. CardioXNet network had the highest accuracy rate (100 percent), whereas recommended bespoke pretrained U-Net model , ResNet-101 had 92.5 percent and 91 percent, respectively.

The total accuracy of the evaluated neural networks revealed that the proposed customized pretrained U-Net model had the highest accuracy of all the networks, with a 94 percent accuracy. Google's Inception V3 network, on the other hand, had the lowest accuracy of the 10 networks tested (71 percent ). Following the sensitivity and specificity results, the suggested customized pre-trained U-Net model with CardioXNet networks were able to reach approx. 94 percent and 93.75 percent accuracy, respectively, whereas the Multipathing method 1 with Segmentation-based method 2 achieved equal accuracy of 91 percent with a 1% difference on specificity (87 percent and 88 percent) and sensitivity (87 percent and 88 percent) (95 percent and 94 percent ).

## V. Conclusion

Chest radiography is a fundamental tool for bulk screening and early identification of lung and heart disorders such as airway obstruction, consolidation, collapse, pleural effusion, and cardiac hyperinflation. Cardiomegaly is indeed a medical disorder characterized by an enlarged heart. Automated methods may be used in a CXR-based screening tool. Because of the creation of new variant deep neural networks (CNNs) models, deep neural networks have become widely used in image processing. The viability of using transfer learning methods for detecting cardiomegaly from X-ray pictures is investigated in this study. In this paper, a supervised neural pre - trained models model for detecting Cardiomegaly illness from X-Ray pictures is proposed. The procedure is carried out in two parts, the first of which involves extracting Cardiomegaly images from the Chest Xray-8 dataset, followed by improving the extracted image quality using an image enhancement approach, and finally compressing the image size. In the second step, a bespoke pertained U-Net model was used to classify the data. The Chest Xray-8 dataset was used to test the use of Python on X-ray images in this study. The proposed customized pertained U-net model obtained 94 percent accuracy, 92.5 percent specificity, and 96.2 percent sensitivity, according to the results. It has also been compared to existing state-of-the-art CNN pertained models, with the proposed model outperforming these models.

The proposed approach can be used to detect additional thoracic disorders in the future, as well. In addition, we intend to collaborate with the medical office to see if our technology can increase the efficiency in manual X-Ray review in clinical settings.